\documentclass[10pt,conference]{IEEEtran}
\usepackage{graphicx}
\usepackage{amsmath}
\usepackage{mathrsfs}
\usepackage{mathtools}
\usepackage{amssymb}
\usepackage{float}
\usepackage{url}
\usepackage{verbatim}
\usepackage{enumerate}
\usepackage{layout}
\usepackage{bm}

\newtheorem{theorem}{Theorem}

\newtheorem{proposition}{Proposition}
\newtheorem{corollary}{Corollary}

\newtheorem{remark}{Remark}

\newcommand{\naturals}{\ensuremath{\mathbb{N}}}

\newcommand{\set}{\ensuremath{\mathcal}}
\newcommand{\OneTo}[1]{[1:#1]}
\newcommand{\eqdef}{\triangleq} % definition
\newcommand{\card}[1]{|#1|}
\newcommand{\bigcard}[1]{\bigl|#1\bigr|}

\RequirePackage{bbm}

\newcommand{\pmfsub}[1]{{\smash{#1}\vphantom{XYZ}}}
\newcommand{\pmf}[1]{\mathsf{#1}}
\newcommand{\pmfOf}[1]{\pmf{P}_{\pmfsub{#1}}}  % the PMF.
\newcommand{\CondpmfOf}[2]{\pmf{P}_{\pmfsub{#1} | \pmfsub{#2}}}
\newcommand{\marginalOf}[2]{\pmf{#1}_{\pmfsub{#2}}}     % marginal of a PMF

\DeclareMathOperator{\Vertex}{\mathsf{V}}
\DeclareMathOperator{\Edge}{\mathsf{E}}
\DeclareMathOperator{\Independentset}{\set{I}}

\DeclareMathOperator{\Degree}{\text{d}}
\newcommand{\V}[1]{\Vertex(#1)}
\newcommand{\E}[1]{\Edge(#1)}
\newcommand{\indset}[1]{\Independentset(#1)}  % independent sets
\newcommand{\dgr}[1]{\Degree_{#1}}   % degree of a vertex

\newcommand{\markov}{\textnormal{\mbox{$\multimap\hspace{-0.73ex}-\hspace{-2ex}-$}}}

\DeclareMathOperator{\Entr}{H}
\DeclareMathOperator{\Info}{I}

\newcommand{\Ent}[1]{\Entr(#1)}

\newcommand{\EntCond}[2]{\Entr(#1 | \kern0.1em #2)}
\newcommand{\bigEntCond}[2]{\Entr\bigl(#1 | \kern0.1em #2\bigr)}
\newcommand{\BigEntCond}[2]{\Entr\Bigl(#1 \kern-0.1em \bigm| \kern-0.1em #2 \Bigr)}
\newcommand{\biggEntCond}[2]{\Entr\biggl(#1 \kern-0.1em \Bigm| \kern-0.1em #2 \biggr)}
\newcommand{\BiggEntCond}[2]{\Entr\Biggl(#1 \kern-0.1em \biggm| \kern-0.1em #2 \Biggr)}

\newcommand{\biggMInfo}[2]{\Info\biggl(#1; \kern0.1em #2 \biggr)}
\newcommand{\BiggMInfo}[2]{\Info\Biggl(#1; \kern0.1em #2 \Biggr)}

\newcommand{\KLDivBin}[2]{\mathsf{D}_{\textnormal{b}}(#1 \kern0.1em \| \kern0.1em #2)} % binary relative entropy
\newcommand{\bigKLDivBin}[2]{\mathsf{D}_{\textnormal{b}}\bigl(#1 \kern0.1em \| \kern0.1em #2 \bigr)}
\newcommand{\BigKLDivBin}[2]{\mathsf{D}_{\textnormal{b}}\Bigl(#1 \kern0.1em \| \kern0.1em #2 \Bigr)}
\newcommand{\biggKLDivBin}[2]{\mathsf{D}_{\textnormal{b}}\biggl(#1 \kern0.1em \| \kern0.1em #2 \biggr)}
\newcommand{\BiggKLDivBin}[2]{\mathsf{D}_{\textnormal{b}}\Biggl(#1 \kern0.1em \| \kern0.1em #2 \Biggr)}
\newcommand{\bigKLDiv}[2]{\mathsf{D}\bigl(#1 \kern0.1em \| \kern0.1em #2 \bigr)}
\newcommand{\BigKLDiv}[2]{\mathsf{D}\Bigl(#1 \kern-0.2em \bigm\| \kern-0.2em #2 \Bigr)}
\newcommand{\biggKLDiv}[2]{\mathsf{D}\biggl(#1 \kern-0.2em \Bigm\| \kern-0.2em #2 \biggr)}
\newcommand{\BiggKLDiv}[2]{\mathsf{D}\Biggl(#1 \kern-0.1em \biggm\| \kern-0.1em #2 \Biggr)}

\begin{document}
\thispagestyle{empty}
\pagestyle{empty}

\title {\Huge{Entropy-Based Proofs of Combinatorial Results on Bipartite Graphs}}
\author{
\IEEEauthorblockN{Igal Sason}
\IEEEauthorblockA{Andrew and Erna Viterbi Faculty of Electrical and Computer Engineering \\
Technion-Israel Institute of Technology, Haifa 3200003, Israel\\
E-mail: sason@ee.technion.ac.il}}

\maketitle

\begin{abstract}
This work considers new entropy-based proofs of some known, or otherwise refined, combinatorial bounds
for bipartite graphs. These include upper bounds on the number of the independent sets, lower bounds on the
minimal number of colors in constrained edge coloring, and lower bounds on the number of walks of a given
length in bipartite graphs. The proofs of these combinatorial results rely on basic properties
of the Shannon entropy.
\end{abstract}

\vspace*{0.2cm}
\section{Introduction}
\label{section: section}

The Shannon entropy serves as a powerful tool in various combinatorial and graph-theoretic
applications (see, e.g., the tutorials in \cite{Galvin14} and \cite{Radhakrishnan01}, as
well as \cite{ErdosR63}, \cite{Friedgut04}, \cite{Massey_IT74}, \cite{Pippenger77},
\cite{Pippenger99}).

Combinatorial properties of bipartite graphs are of great interest in graph theory,
combinatorics, modern coding theory, and information theory. Entropy-based
proofs pertain to the following aspects of such combinatorial properties:
\begin{enumerate}[1)]
\item {\em Enumeration of the independent sets in graphs}:
Many important structures can be modeled by independent sets in a graph, i.e., subsets of vertices
in a graph where none of them are connected by an edge. If a graph models some kind
of incompatibility, then an independent set in this graph represents a mutually compatible collection.
An application of Shearer's lemma to obtain a tight upper bound on the number of independent
sets in $d$-regular bipartite graphs, and a recent extension of this information-theoretic proof for
irregular bipartite graphs (which is tight for bipartite graphs that are regular on one side)
are available in \cite{Kahn01} and \cite{Sason21}, respectively. The information-theoretic literature
considers independent sets and their enumeration in \cite{JohnsonKM13}, \cite{Kahn01}, \cite{MadimanT_ISIT07},
\cite{MadimanT_IT10}, and recently in \cite{Sason21}.

\item {\em Enumeration of perfect matchings in bipartite graphs}: An elegant entropy-based proof
of Bregman's theorem, which is a tight upper bound on the permanent of square matrices with binary entries,
was introduced in \cite{Radhakrishnan97}. The permanent of such a matrix is equivalent to the number
of perfect matchings of the induced bipartite graph, so \cite{Radhakrishnan97} provides an information-theoretic
proof for a tight upper bound on the number of perfect matchings in bipartite graphs.

\item {\em Moore's bound}: The girth of a graph is the length of its shortest cycle, and it is of interest
in graph theory. The girth is also of importance in the realm of modern coding theory with respect to codes
defined on (bipartite) graphs and their iterative message-passing decoding algorithms
(see, e.g., \cite{Gallager63} and \cite[Problems~3.25--3.37]{MCT}). Moore's bound provides an upper bound
on the girth of irregular graphs as a function of their number of vertices and their average degree \cite{AlonHL02}.
An entropy-based proof of Moore's bound for graphs (and bipartite graphs) was introduced in \cite{BabuR14}.

\item {\em Additional combinatorial properties of bipartite graphs}: In \cite{KacedRV18}, a new conditional
entropy inequality was derived, followed by a study of two of its combinatorial applications to bipartite
graphs. These include a derivation of a lower bound on the minimal number of colors in (rich) graph coloring,
and a derivation of a lower bound on the biclique cover number of bipartite graphs.

\item Although not directly related to bipartite graphs, a variant of
Shearer's lemma for the relative entropy was introduced in \cite{MadimanT_IT10} (see
Corollary~7 and Remark~9 there), and also independently (several years later) in \cite{GavinkyLSS14}.
Furthermore, the work in \cite{GavinkyLSS14} applies this variant to obtain a Chernoff-type bound for the sum of
read-$k$ functions of independent variables (i.e., a set of functions where each variable participates
in at most $k$ functions); it is then used in \cite{GavinkyLSS14} to derive a probabilistic bound on the number of triangles
in random graphs, constructed by the Erd\H{o}s-R\'{e}nyi model.
\end{enumerate}

There is a long history of applying entropy inequalities for obtaining combinatorial results, and the present
paper aims to further develop this connection.
It provides new entropy-based proofs of known, or otherwise refined, combinatorial bounds
for bipartite graphs. This paper has the following structure: Section~\ref{section: preliminaries and notation}
provides preliminaries and notation, and Sections~\ref{section: number of independent sets}--\ref{section: number of walks}
suggest entropy-based proofs of combinatorial properties of bipartite graphs. Specifically,
Section~\ref{section: number of independent sets} refers to a generalized information-theoretic approach for bounding
the number of independent sets in bipartite graphs.
The material in Section~\ref{section: number of independent sets} outlines our recent work in \cite{Sason21}.
Sections~\ref{section: edge coloring} and~\ref{section: number of walks} provide entropy-based
proofs of two combinatorial results for bipartite graphs.
Section~\ref{section: edge coloring} generalizes a conditional-entropy inequality in \cite{KacedRV18}
(Proposition~\ref{prop: generalized cond. ineq.} here). In continuation to
\cite[Section~IV]{KacedRV18}, the generalized inequality is used to derive a lower bound on the minimal
number of colors in constrained graph colorings of bipartite graphs.  Section~\ref{section: number of walks}
provides entropy-based lower bounds on the number of walks of a given length in bipartite graphs
(Proposition~\ref{proposition: number of walks}), relying on a work
on the Moore bound \cite{AlonHL02}, and its later information-theoretic
formulation in \cite{BabuR14}.

\section{Preliminaries and Notation}
\label{section: preliminaries and notation}

Let $G$ be an undirected graph, and let $\V{G}$ and $\E{G}$ denote, respectively, the sets
of vertices and edges in $G$.

A graph is called {\em bipartite} if it has two types of vertices, and an edge cannot
connect vertices of the same type; we refer to the vertices of a bipartite graph $G$
as left and right vertices.

A graph $G$ is called {\em complete} if every vertex $v \in \V{G}$ is connected to all the
other vertices in $\V{G} \setminus \{v\}$ (and not to itself); similarly, a bipartite graph
is called complete if every vertex is connected to all the vertices of the other type in
the graph. A complete $(d-1)$-regular graph is denoted by $K_d$, having a number of vertices
$\bigcard{\V{K_d}} = d$, and a number of edges $\bigcard{\E{K_d}} = \tfrac12 \, d(d-1)$.
Likewise, a complete $d$-regular bipartite graph is denoted by $K_{d,d}$, having a number
of vertices $\bigcard{\V{K_{d,d}}} = 2d$ (i.e., $d$ vertices of each of the two types),
and a number of edges $\bigcard{\E{K_{d,d}}} = d^2$.

An {\em independent set} of an undirected graph $G$ is a subset of its vertices such that
none of the vertices in this subset are adjacent (i.e., none of them are joined by an edge).
Let $\indset{G}$ denote the set of all the independent sets in $G$, and let $\bigcard{\indset{G}}$
denote the number of independent sets in $G$.

The {\em tensor product} $G \times H$ of two graphs $G$ and $H$ is a graph
such that the vertex set of $G \times H$ is the Cartesian product $\V{G} \times \V{H}$,
and two vertices $(g,h), (g', h') \in \V{G \times H}$ are adjacent if and only if
$g$ is adjacent to $g'$, and $h$ is adjacent to $h'$ (i.e., $(g,g') \in \E{G}$ and
$(h,h') \in \E{H}$).

By the definition of a complete $d$-regular graph $K_d$, the graph $K_2$
is specialized to two vertices that are connected by an edge. Let us label the
two vertices in $K_2$ by 0 and~1. For a graph $G$, the tensor product
$G \times K_2$ is a bipartite graph, called the {\em bipartite double cover}
of $G$, where the set of vertices in $G \times K_2$ is given by
\begin{IEEEeqnarray}{rcl}
\label{eq: vertices of bipartite double cover}
\V{G \times K_2} = \bigl\{ (v,i) : v \in \V{G}, \, i \in \{0,1\} \bigr\},
\end{IEEEeqnarray}
and its set of edges is given by
\begin{IEEEeqnarray}{rCl}
\label{eq: edges of bipartite double cover}
\E{G \times K_2} = \bigl\{ \bigl( (u,0), (v,1) \bigr) : (u,v) \in \E{G} \bigr\}.
\end{IEEEeqnarray}
Every edge $e = (u,v) \in \E{G}$ is mapped into the two edges
$\bigl( (u,0), (v,1) \bigr) \in \E{G \times K_2}$ and
$\bigl( (v,0), (u,1) \bigr) \in \E{G \times K_2}$
(since the graph $G$ is undirected). This implies that the numbers of vertices
and edges in $G \times K_2$ are doubled in comparison to their respective numbers
in $G$; moreover, every edge in $G$, which connects a pair of
vertices of specified degrees, is mapped into two edges in $G \times K_2$ where
each of these two edges connects a pair of vertices of the same specified degrees.

An {\em edge coloring} of a graph is an assignment of colors to its edges
such that each two edges sharing a vertex in this graph have different colors. Finding the minimal
number of colors in an edge coloring of a given graph is a classical problem in
graph theory (see, e.g., \cite[Chapter~5]{Diestel}).

The following further notation and basic facts are used:
\begin{itemize}
\item $\naturals \eqdef \{1, 2, \ldots \}$ denotes the set of natural numbers.
\item $X^n \eqdef (X_1, \ldots X_n)$ denotes an $n$-dimensional random vector
of discrete random variables, having a joint probability mass function (PMF)
that is denoted by $\pmfOf{X^n}$.
\item For every $n \in \naturals$, let $\OneTo{n} \eqdef \{1, \ldots, n\}$;
\item $X_{\set{S}} \eqdef (X_i)_{i \in \set{S}}$ is a random vector for
an arbitrary nonempty subset $\set{S} \subseteq \OneTo{n}$; if
$\set{S} = \emptyset$, then conditioning on $X_{\set{S}}$ is void.
%\item Let $X$ be a discrete random variable that takes its values on a set $\set{X}$,
%and let $\pmfOf{X}$ be the PMF of $X$. The {\em Shannon entropy} of $X$ is given by
%\begin{IEEEeqnarray}{rCl}
%\label{eq: entropy}
%\Ent{X} \eqdef -\sum_{x \in \set{X}} \pmfOf{X}(x) \, \log \pmfOf{X}(x),
%\end{IEEEeqnarray}
%where we take all logarithms to base~2.
%\item For $p \in [0,1]$,
%\begin{IEEEeqnarray}{rCl}
%\label{eq2: EntBin}
%\EntBin{p} \eqdef -p \log p -(1-p) \log(1-p),
%\end{IEEEeqnarray}
%where $\EntBin{\cdot}$ is the {\em binary entropy function}.
By continuous extension, the convention $0 \log 0 = 0$ is used.
%\item If $\set{X}$ is a finite set, then
%$\Ent{X} \leq \log \card{\set{X}}$,
%with equality if and only if $X$ is equiprobable over $\set{X}$.
%\item Let $X$ and $Y$ be discrete random variables with a joint PMF $\pmfOf{XY}$, and having a conditional
%PMF of $X$ given $Y$ which is denoted by $\CondpmfOf{X}{Y}$. Let $X$ and $Y$ take their values in the
%sets $\set{X}$ and $\set{Y}$, respectively. The {\em conditional entropy} of $X$ given $Y$
%is defined as
%\begin{IEEEeqnarray}{rCl}
%\label{eq: conditional entropy 1}
%\EntCond{X}{Y} & \eqdef & -\sum_{(x,y) \in \set{X} \times \set{Y}} \pmfOf{XY}(x,y) \log \CondpmfOf{X}{Y}(x|y) \\
%\label{eq: conditional entropy 1.5}
%&=& \sum_{y \in \set{Y}} \pmfOf{Y}(y) \, \EntCond{X}{Y=y}.
%\end{IEEEeqnarray}
%\item Conditioning cannot increase the entropy, i.e.,
%\begin{align}
%\label{eq: conditioning reduces entropy}
%\EntCond{X}{Y} \leq \Ent{X},
%\end{align}
%with equality if and only if $X$ and $Y$ are independent.
%\item The chain rule for the Shannon entropy is given by
%\begin{align}
%\label{eq2: chain rule}
%\Ent{X^n} &= \sum_{i=1}^n \EntCond{X_i}{X^{i-1}}.
%\end{align}
\end{itemize}

Shearer's lemma extends the subadditivity property of the Shannon entropy.
\begin{proposition}[Shearer's Lemma, \em{\cite{ChungGFS86}}]
\label{proposition: Shearer's Lemma}
Let $X_1, \ldots, X_n$ be discrete random variables, and let
$\set{S}_1, \ldots, \set{S}_m \subseteq \OneTo{n}$ include every
element $i \in \OneTo{n}$ in {\em at least} $k \geq 1$ of these
subsets. Then,
\begin{IEEEeqnarray}{rCl}
\label{eq: Shearer's Lemma}
k \Ent{X^n} \leq \sum_{j=1}^m \Ent{X_{\set{S}_j}}.
\end{IEEEeqnarray}
\end{proposition}

\begin{remark}[\cite{Sason21}]
\label{remark: Sheaerer's lemma}
Inequality \eqref{eq: Shearer's Lemma} holds even if the sets $\set{S}_1, \ldots, \set{S}_m$ are
not necessarily included in $\OneTo{n}$. To verify it, define
$\set{S}'_j \eqdef \set{S}_j \cap [1:n]$ for $j \in \OneTo{m}$. The subsets
$\set{S}'_1, \ldots, \set{S}'_m$ are included in $\OneTo{n}$, and every element
$i \in \OneTo{n}$ continues to be included in at least $k \geq 1$ of these subsets. Hence,
Proposition~\ref{proposition: Shearer's Lemma} can be applied to the subsets
$\set{S}'_1, \ldots, \set{S}'_m$. By the monotonicity property of the entropy, the inclusion
$\set{S}'_j \subseteq \set{S}_j$ implies that $\Ent{X_{\set{S}'_j}} \leq \Ent{X_{\set{S}_j}}$
for all $j \in \OneTo{m}$, which then yields the satisfiability of \eqref{eq: Shearer's Lemma}.
\end{remark}

\section{Number of Independent Sets}
\label{section: number of independent sets}

\subsection{Background}
This present section is focused on the problem of upper bounding the number of independent sets in a graph,
expressed in terms of its degree distribution. For bipartite regular graphs, Kahn \cite{Kahn01} established
a tight upper bound using an information-theoretic approach, which is given as follows.

\begin{theorem}[Kahn 2001, \cite{Kahn01}]
\label{theorem: Kahn 2001}
If $G$ is a bipartite $d$-regular graph with $n$ vertices, then
\begin{IEEEeqnarray}{rCl}
\label{eq: Kahn 2001}
\bigcard{\indset{G}} \leq \bigl( 2^{d+1} - 1 \bigr)^{\frac{n}{2d}}.
\end{IEEEeqnarray}
Moreover, if $n$ is an even multiple of $d$, then the upper bound in
the right side of \eqref{eq: Kahn 2001} is tight, and it is obtained
by a disjoint union of $\frac{n}{2d}$ complete $d$-regular bipartite
graphs $(K_{d,d})$.
\end{theorem}

Kahn also conjectured in \cite{Kahn01} an upper bound for general graphs. His conjectured bound was recently
proved (after two decades) by Sah et al. (2019), using different techniques not involving information theory.
Their tight bound is as follows:
\begin{theorem}[Sah et al. 2019, \cite{SahSaStZhao19}]
\label{theorem: Sah et al., 2019}
Let $G$ be an undirected graph without isolated vertices or multiple edges
connecting any pair of vertices. Let $\dgr{v}$ denote the degree of a vertex
$v \in \V{G}$. Then,
\begin{IEEEeqnarray}{rCl}
\label{eq: Sah et al.}
\bigcard{\indset{G}} \leq \prod_{(u,v) \in \E{G}}
(2^{\dgr{u}} + 2^{\dgr{v}} - 1)^{\frac1{\dgr{u} \dgr{v}}}
\end{IEEEeqnarray}
with equality if $G$ is a disjoint union of complete bipartite graphs.
\end{theorem}

The main contribution of our recent work in \cite{Sason21} is the extension of
Kahn’s information-theoretic proof technique to handle irregular bipartite graphs. In particular, when
the bipartite graph is regular on one side, but it may be irregular in the other, the extended entropy-based
proof technique yields the same bound that was conjectured by Kahn \cite{Kahn01} and proved by Sah et al. \cite{SahSaStZhao19}.

The following result by Zhao \cite{Zhao10} upper bounds the square of the number
of independent sets of an arbitrary finite graph $G$ by the number of independent
sets of the bipartite double cover of this graph (i.e., the tensor product of $G$
with $K_2$).
\begin{theorem}[Zhao 2010, \cite{Zhao10}]
\label{theorem: Zhao10}
For every finite graph $G$:
\begin{IEEEeqnarray}{rCl}
\label{eq: Zhao's inequality}
\bigcard{\indset{G}}^2 \leq \bigcard{\indset{G \times K_2}}.
\end{IEEEeqnarray}
\end{theorem}
As an application of Theorem~\ref{theorem: Zhao10}, the extension of \eqref{eq: Sah et al.} from
bipartite graphs to general graphs (without isolated vertices or multiple edges) was enabled in
\cite[Lemma~3]{GalvinZ11} by relying on \eqref{eq: Zhao's inequality}. Recall that every edge
$e = (u,v) \in \E{G}$, which is connected in a graph $G$ to a pair of vertices of degrees $\dgr{u}$
and $\dgr{v}$, is mapped into two edges in the bipartite graph $G \times K_2$ where each one of
these edges is connected to a pair of vertices of degrees $\dgr{u}$ and $\dgr{v}$ (see
Section~\ref{section: preliminaries and notation}). By inequality \eqref{eq: Zhao's inequality},
and in view of the above proof for the setting of bipartite graphs, we obtain that for a general
graph $G$
\begin{IEEEeqnarray}{rCl}
\bigcard{\indset{G}}^2 & \leq & \bigcard{\indset{G \times K_2}} \label{eq5: UB cardinality}  \\
& \leq & \prod_{(u',v') \in \E{G \times K_2}} \bigl( 2^{\dgr{u'}}
+ 2^{\dgr{v'}} - 1 \bigr)^{\frac1{\dgr{u'} \, \dgr{v'}}} \label{eq6: UB cardinality}  \\
&=& \prod_{(u,v) \in \E{G}} \bigl( 2^{\dgr{u}} + 2^{\dgr{v}}
- 1 \bigr)^{\frac2{\dgr{u} \, \dgr{v}}}.  \label{eq7: UB cardinality}
\end{IEEEeqnarray}
Finally, taking the square-roots of the left side of \eqref{eq5: UB cardinality} and the right side of
\eqref{eq7: UB cardinality} gives \eqref{eq: Sah et al.} for general graphs.

In view of the above paragraph, it is sufficient to prove Theorem~\ref{theorem: Sah et al., 2019}
for general graphs by confirming it in the special setting of bipartite graphs.
The work in \cite{SahSaStZhao19} recently proved Theorem~\ref{theorem: Sah et al., 2019} for bipartite graphs
by using an induction on the number of vertices in a graph $G$, and by obtaining a recurrence inequality whose
derivation involves judicious applications of H\"{o}lder's inequality (see
\cite[Sections~2 and 4]{SahSaStZhao19}). The proof there does not rely on information theory.

\subsection{Contribution}
In a very recent paper \cite{Sason21}, we provide an extension of the entropy-based proof by Kahn \cite{Kahn01}
from bipartite $d$-regular graphs to general bipartite graphs, and then we prove \eqref{eq: Sah et al.}
for the family of bipartite graphs that are regular on one side (see \cite[Section~4]{Sason21}).
The proof in \cite{Sason21} follows the same recipe of Kahn's proof in \cite{Kahn01} with
some complications that arise from the non-regularity of the bipartite graphs. The proof in \cite{Sason21}
deviates from the proof in \cite{Kahn01} already at its starting point, by a proper adaptation
of the proof technique to the general setting of irregular bipartite graphs, followed by a bit
more complicated usage of Shearer's lemma (in light of Remark~\ref{remark: Sheaerer's lemma})
and a more involved analysis. The reader is referred to the proof in \cite[Section~4]{Sason21}.

In \cite[Section~5]{Sason21}, we suggest a variant of the proof of Zhao's Inequality in
\eqref{eq: Zhao's inequality} (given implicitly in \cite[Lemma~2.1]{Zhao10}, and
explicitly in a follow-up work by Galvin and Zhao \cite{GalvinZ11}).
This forms in essence a reformulation of Zhao’s proof, which is provided as follows.

\begin{IEEEproof}
Let $G$ be a finite graph, and let $\bigcard{\V{G}} = n$. Label the vertices in the left and
right sides of the bipartite graph $G \times K_2$ (i.e., the bipartite double cover of $G$)
by $\{(i,0)\}_{i=1}^n$ and $\{(i,1)\}_{i=1}^n$, respectively.

Choose independently and uniformly at random two independent sets $\set{S}_0, \set{S}_1 \in \indset{G}$.
For $i \in \OneTo{n}$, let $X_i, Y_i \in \{0,1\}$ be random variables defined as $X_i = 1$
if and only if $i \in \set{S}_0$, and $Y_i = 1$ if and only if $i \in \set{S}_1$. Then, by the
statistical independence and equiprobable selection of the two independent sets from $\indset{G}$,
we have
\begin{IEEEeqnarray}{rCl}
\Ent{X^n, Y^n} &=& \Ent{X^n} + \Ent{Y^n} \label{eq: Ent38} \\
&=& 2 \log \, \bigcard{\indset{G}},   \label{eq: Ent39}
\end{IEEEeqnarray}
where \eqref{eq: Ent38} holds since $X^n = (X_1, \ldots, X_n)$ and $Y^n = (Y_1, \ldots, Y_n)$ are
statistically independent (by construction), and \eqref{eq: Ent39} holds since they both have an
equiprobable distribution over a set whose cardinality is $\bigcard{\indset{G}}$.

Consider the following set of vertices in $G \times K_2$:
\begin{IEEEeqnarray}{rCl}
\label{eq: set S}
\set{S} &\eqdef& \bigl\{ \set{S}_0 \times \{0\} \bigr\} \bigcup \bigl\{ \set{S}_1 \times \{1\} \bigr\} \\[0.1cm]
&=& \bigcup_{i \in \set{S}_0, \, j \in \set{S}_1} \bigl\{ (i,0), (j,1) \bigr\}.
\end{IEEEeqnarray}
The set $\set{S}$ is not necessarily an independent set in $G \times K_2$ since $\bigl( (i,0), (j,1) \bigr) \in \E{G \times K_2}$
for all $i \in \set{S}_0$ and $j \in \set{S}_1$ for which $(i,j) \in \E{G}$ (see \eqref{eq: edges of bipartite double cover}).
We next consider all $(i,j) \in \E{G}$ such that $X_i = Y_j = 1$. To that end, fix an ordering of all the $2^n$ subsets of $\V{G}$,
and let $\set{T} \in \V{G}$ be the first subset in this particular ordering that includes exactly one endpoint of each edge
$(i,j) \in \E{G}$ for which $X_i = Y_j = 1$. Consider the following replacements:
\begin{itemize}
\item If $(i,0) \in \set{S}$ and $i \in \set{T}$, then $(i,0)$ is replaced by $(i,1)$;
\item Likewise, if $(j,1) \in \set{S}$ and $j \in \set{T}$, then $(j,1)$ is replaced by $(j,0)$.
\end{itemize}
Let $\widetilde{\set{S}}$ be the set of new vertices after these possible replacements.
Then, $\widetilde{S} \in \indset{G \times K_2}$ since all adjacent vertices in $\set{S}$ are no longer connected in $\widetilde{\set{S}}$.
Indeed, there is no way that after (say) a vertex $(i,0)$ is replaced by $(i,1)$, there is another replacement
of a vertex $(j,1)$ by $(j,0)$, for some $j$ such that $(i,j) \in \E{G}$; otherwise, that would
mean that $\set{T}$ contains both $i$ and $j$, which is impossible by construction.

Similarly to the way $X^n, Y^n \in \{0,1\}^n$ were defined, let $\widetilde{X}^n, \widetilde{Y}^n \in \{0,1\}^n$ be defined such that,
for all $i \in \OneTo{n}$, $\widetilde{X}_i = 1$ if and only if $(i,0) \in \widetilde{\set{S}}$, and $\widetilde{Y}_i = 1$ if and only if
$(i,1) \in \widetilde{\set{S}}$.

The mapping from $(X^n,Y^n)$ to $(\widetilde{X}^n, \widetilde{Y}^n)$ is injective. Indeed, it is shown to be injective by finding all
indices $(i,j) \in \E{G}$ such that $\widetilde{X}_i = \widetilde{X}_j = 1$ or $\widetilde{Y}_i = \widetilde{Y}_j = 1$, finding the first
subset $\set{T} \in \V{G}$ according to our previous fixed ordering of the $2^n$ subsets of $\V{G}$ that includes exactly one endpoint
of each such edge $(i,j) \in \E{G}$, and performing the reverse operation to return back to $X^n$ and $Y^n$ (e.g., if $(i,j) \in \E{G}$,
$\widetilde{X}_i = \widetilde{X}_j = 1$ and $i \in \set{T}$ while $j \not\in \set{T}$, then $\widetilde{X}_i = 1$ is transformed back
to $Y_i = 1$, and $\widetilde{X}_j = 1$ is transformed back to $X_j = 1$). Consequently, we get
\begin{IEEEeqnarray}{rCl}
\Ent{X^n, Y^n} &=& \Ent{\widetilde{X}^n, \widetilde{Y}^n} \label{eq: Ent40} \\
& \leq & \log \, \bigcard{\indset{G \times K_2}},  \label{eq: Ent41}
\end{IEEEeqnarray}
where \eqref{eq: Ent40} holds by the injectivity of the mapping from $(X^n, Y^n)$ to $(\widetilde{X}^n, \widetilde{Y}^n)$,
and \eqref{eq: Ent41} holds since $\widetilde{S}$ is an independent set in $G \times K_2$, which implies that
$(\widetilde{X}^n, \widetilde{Y}^n)$ can get at most $\bigcard{\indset{G \times K_2}}$ possible values (by definition,
there is a one-to-one correspondence between $\widetilde{\set{S}}$ and $(\widetilde{X}^n, \widetilde{Y}^n)$).
Combining \eqref{eq: Ent38}, \eqref{eq: Ent39}, \eqref{eq: Ent40} and \eqref{eq: Ent41} gives
\begin{IEEEeqnarray}{rCl}
\label{eq: final}
2 \log \, \bigcard{\indset{G}} \leq \log \, \bigcard{\indset{G \times K_2}},
\end{IEEEeqnarray}
which gives \eqref{eq: Zhao's inequality} by exponentiation of both sides of \eqref{eq: final}.
\end{IEEEproof}

\subsection{Outlook}
In \cite[Corollary~6.2]{JohnsonKM13} and its related discussion, Johnson, Kontoyiannis and Madiman
provided an upper bound on the entropy of the size of a random independent set in a claw-free graph
(i.e., a graph that does not contain the complete bipartite graph $K_{1,3}$ as an induced subgraph).
In light of a connection between the size of a random independent set and
the total number of independent sets (the latter is the partition function of the hard-core model
with fugacity~1, see the definition of the independence polynomial in \cite{Zhao10} for details),
it is left for future work to study if the results in \cite{Sason21} can be applied to yield bounds on
the size of a random independent set, or bounds on the partition function with a general fugacity.

\section{Edge Coloring of Bipartite Graphs}
\label{section: edge coloring}

In \cite[Theorem~1]{KacedRV18}, the following result is proved.
\begin{theorem}[Kaced et al. 2018, \cite{KacedRV18}]
\label{theorem: Kaced}
Let $A, X$ and $Y$ be discrete random variables taking their values
in the sets $\set{A}, \set{X}, \set{Y}$, respectively, with
a joint probability mass function $\pmf{P}_{A,X,Y}$.
If for every $(x,y) \in \set{X} \times \set{Y}$,
there exists {\em at most one element} $a \in \set{A}$ such that
$\marginalOf{P}{A,X}(a,x) \, \marginalOf{P}{A,Y}(a,y) > 0$
then
\begin{IEEEeqnarray}{rCl}
\EntCond{A}{X} + \EntCond{A}{Y} \leq \Ent{A}.
\label{eq: Kaced}
\end{IEEEeqnarray}
\end{theorem}

We next provide a modest generalization of Theorem~\ref{theorem: Kaced}, which
suggests an extension of the result in \cite[Corollary~1]{KacedRV18} with
respect to edge coloring of bipartite graphs.

\begin{proposition}
\label{prop: generalized cond. ineq.}
Let $A, X, Y$ be discrete random variables taking values
in sets $\set{A}, \set{X}, \set{Y}$, respectively. Then,
\begin{IEEEeqnarray}{rCl}
\EntCond{A}{X} + \EntCond{A}{Y} \leq \Ent{A} + \log m,
\label{eq: generalized cond. ineq.}
\end{IEEEeqnarray}
where
\begin{IEEEeqnarray}{rCl}
\hspace*{-0.5cm} m \eqdef \sup_{(x,y) \in \set{X} \times \set{Y}}
\bigcard{ \bigl\{a \in \set{A}: \,  \marginalOf{P}{A,X}(a,x) \,
\marginalOf{P}{A,Y}(a,y) > 0 \bigr\} }. \label{eqdef: m}
\end{IEEEeqnarray}
\end{proposition}

\begin{IEEEproof}
Consider the PMF
\begin{IEEEeqnarray*}{rCl}
\marginalOf{P'}{A,X,Y}(a,x,y) \eqdef
\begin{dcases}
\frac{\marginalOf{P}{A,X}(a,x) \, \marginalOf{P}{A,Y}(a,y)}{\marginalOf{P}{A}(a)}, & \marginalOf{P}{A}(a) > 0 \\
0, & \text{otherwise,}
\end{dcases}
\end{IEEEeqnarray*}
which refers to the case where $A$ is chosen according to the PMF $\marginalOf{P}{A}$, and
$X$ and $Y$ are conditionally independent given $A$ with the conditional
PMFs $\CondpmfOf{X}{A}$ and $\CondpmfOf{Y}{A}$, respectively.
\begin{IEEEeqnarray}{rCl}
&& \hspace*{-0.3cm} \EntCond{A}{X} + \EntCond{A}{Y} - \Ent{A} \nonumber \\[0.1cm]
&& = \Ent{A,X} + \Ent{A,Y} - \Ent{X} - \Ent{Y} - \Ent{A} \label{200121a} \\[0.1cm]
&& = \sum_{a,x,y} \marginalOf{P}{A,X,Y}(a,x,y) \,
\log \left( \frac{\marginalOf{P}{X}(x) \, \marginalOf{P}{Y}(y) \,
\marginalOf{P}{A}(a)}{\marginalOf{P}{A,X}(a,x) \marginalOf{P}{A,Y}(a,y)} \right) \nonumber \\[0.1cm]
&& = \sum_{a,x,y} \marginalOf{P'}{A,X,Y}(a,x,y) \,
\log \left( \frac{\marginalOf{P}{X}(x) \, \marginalOf{P}{Y}(y) \,
\marginalOf{P}{A}(a)}{\marginalOf{P}{A,X}(a,x) \, \marginalOf{P}{A,Y}(a,y)} \right) \nonumber
\end{IEEEeqnarray}
where the last equality holds since the following equalities are satisfied by the marginals
of $\marginalOf{P}{A,X,Y}$ and $\marginalOf{P'}{A,X,Y}$:
\begin{IEEEeqnarray}{rCl}  \label{200121b}
\marginalOf{P}{A,X} = \marginalOf{P'}{A,X}, \quad \marginalOf{P}{A,Y} = \marginalOf{P'}{A,Y},
\end{IEEEeqnarray}
and the terms $\Ent{A,X}, \Ent{A,Y}, \Ent{X}, \Ent{Y}, \Ent{A}$ in the right-side
of \eqref{200121a} only depend on the marginal PMFs that appear in \eqref{200121b}.
Due to the concavity of the logarithmic function, invoking Jensen's inequality gives
\begin{IEEEeqnarray}{rCl}
&& \hspace*{-0.2cm} \EntCond{A}{X} + \EntCond{A}{Y} - \Ent{A} \nonumber \\[0.1cm]
&& \leq \log \sum_{(a,x,y) \in \text{supp}(\pmf{P'})}
\frac{\marginalOf{P'}{A,X,Y}(a,x,y) \, \marginalOf{P}{X}(x) \, \marginalOf{P}{Y}(y) \,
\marginalOf{P}{A}(a)}{\marginalOf{P}{A,X}(a,x) \, \marginalOf{P}{A,Y}(a,y)} \nonumber \\
&& = \log \left( \sum_{(a,x,y) : \marginalOf{P'}{A,X,Y}(a,x,y) > 0}
\marginalOf{P}{X}(x) \, \marginalOf{P}{Y}(y) \right).  \label{200121c}
\end{IEEEeqnarray}
For all $(x,y) \in \set{X} \times \set{Y}$, let
\begin{IEEEeqnarray}{rCl}
m(x,y) &&\eqdef \bigcard{ \bigl\{a \in \set{A}: \,  \marginalOf{P}{A,X}(a,x) \,
\marginalOf{P}{A,Y}(a,y) > 0 \bigr\} } \label{200121d} \\[0.1cm]
&&= \bigcard{ \bigl\{a \in \set{A}: \, \marginalOf{P'}{A,X,Y}(a,x,y) > 0 \bigr\} }.  \label{200121e}
\end{IEEEeqnarray}
Then,
\begin{IEEEeqnarray}{rCl}
&& \hspace*{-1.5cm} \sum_{(a,x,y) : \marginalOf{P'}{A,X,Y}(a,x,y) > 0}
\marginalOf{P}{X}(x) \, \marginalOf{P}{Y}(y) \nonumber \\
&&= \sum_{x,y} \bigl\{ m(x,y) \, \marginalOf{P}{X}(x) \, \marginalOf{P}{Y}(y) \bigr\} \label{200121f} \\
&&\leq m \sum_{x,y} \bigl\{ \marginalOf{P}{X}(x) \, \marginalOf{P}{Y}(y) \bigr\} \label{200121g} \\
&&=m, \label{200121h}
\end{IEEEeqnarray}
where \eqref{200121f} holds by \eqref{200121d}, and \eqref{200121g} holds by \eqref{eqdef: m}.
Finally, combining \eqref{200121c} and \eqref{200121f}--\eqref{200121h} gives
\eqref{eq: generalized cond. ineq.}.
\end{IEEEproof}

Proposition~\ref{prop: generalized cond. ineq.} is useful if $\log m < \Ent{A}$, and
otherwise \eqref{eq: generalized cond. ineq.} becomes trivial.

In view of Proposition~\ref{prop: generalized cond. ineq.} and the proof of
\cite[Corollary~1]{KacedRV18}, the following result follows readily.
\begin{corollary}
\label{corollary: gen. Kaced et al.}
Consider a bipartite graph $G$ with minimal left and right degrees that are equal to
$d_{\text{L}}$ and $d_{\text{R}}$, respectively. Consider an edge coloring of $G$
where, in addition to the requirement that every two edges sharing a node have
different colors, it is satisfied that
for all pairs of left node $v_{\text{L}}$ and right node $v_{\text{R}}$ in $\V{G}$,
there are at most $m$ colors touching both $v_{\text{L}}$ and $v_{\text{R}}$.
Then, the number of colors in every such edge coloring of $G$ is at least
$\frac{d_{\text{L}} \, d_{\text{R}}}{m}$.
\end{corollary}
\begin{IEEEproof}
It is similar to the proof of \cite[Corollary~1]{KacedRV18}, with the only modification
of using \eqref{eq: generalized cond. ineq.} with an arbitrary $m \in \naturals$ (instead
of \eqref{eq: Kaced}, referring the special case where $m=1$).
\end{IEEEproof}

\vspace*{0.1cm}
By Vizing's theorem on edge coloring of graphs \cite{Vizing64} (see \cite[Theorem~5.3.2]{Diestel}),
the number of colors needed to edge color a simple graph (i.e., an undirected graph containing no
graph loops or multiple edges) is either equal to its maximal degree $d_{\max}$ or to $d_{\max}+1$.
Furthermore, for bipartite graphs, the number of colors is always equal to $d_{\max}$.

The following simple consequence of Corollary~\ref{corollary: gen. Kaced et al.} motivates the
extension of the result in \cite[Corollary~1]{KacedRV18} for constrained edge colorings of
bipartite graphs. It provides a refinement of Vizing's theorem for regular bipartite graphs.
\begin{corollary}
If $G$ is a $d$-regular bipartite graph, and there is a requirement on the richness of the colors
in the sense that at most $m<d$ colors touch every pair of left and right vertices in $G$, then
the number of required colors is at least $\frac{d^2}{m}$
(which is strictly larger than $d_{\max} = d$).
\end{corollary}

\section{Number of Walks of a Given Length}
\label{section: number of walks}

The present section derives lower bounds on the number of walks
of a given length in bipartite graphs, based on basic properties of
the Shannon entropy. These results rely on the work by Alon, Hoory
and Linial on the Moore bound \cite{AlonHL02}, and its later information-theoretic
formulation due to Babu and Radhakrishnan \cite{BabuR14}.

We introduce here the refined bounds in \eqref{eq: LB1, k odd} and \eqref{eq: LB1, k even},
which are expressed in terms of Shannon entropies of probability mass functions that
are induced by the degree distributions of the bipartite graph; these lower bounds
tighten the bounds in \eqref{eq: LB2, k odd} and \eqref{eq: LB2, k even}, respectively.
\begin{proposition}
\label{proposition: number of walks}
Let $G$ be a bipartite graph with a disjoint partition of its
vertex set $\V{G}$ to sets of left and right vertices $\set{U}$ and
$\set{V}$, respectively, with $\card{\set{U}} = m$ and $\card{\set{V}} = n$.
Let $\set{P}_k$ be the set of all walks of a given length $k \in \naturals$,
where edges may be repeated. Let $\dgr{r}$ denote the degree of a vertex
$r \in \V{G}$, and let $\pmf{P}$ and $\pmf{Q}$ be PMFs defined, respectively,
on $\set{U}$ and $\set{V}$ as follows:
\begin{IEEEeqnarray}{rCl}
\label{eq: PMF P - left side}
&& \pmf{P}(u) \eqdef \frac{\dgr{u}}{\card{\E{G}}}, \quad u \in \set{U}, \\[0.1cm]
\label{eq: PMF Q - right side}
&& \pmf{Q}(v) \eqdef \frac{\dgr{v}}{\card{\E{G}}}, \quad v \in \set{V}.
\end{IEEEeqnarray}
\begin{enumerate}[1)]
\item If $k$ is odd, then
\begin{IEEEeqnarray}{rCl}
\label{eq: LB1, k odd}
\hspace*{-0.3cm} \bigcard{ \set{P}_k } &\geq& \card{\E{G}}^k \,
\exp \bigl( -\tfrac12 (k-1) [ \Ent{P} + \Ent{Q} ] \bigr) \\[0.2cm]
\label{eq: LB2, k odd}
&\geq& \frac{\card{\E{G}}^k}{(mn)^{\frac{k-1}{2}}}.
\end{IEEEeqnarray}
\item If $k$ is even, then
\begin{IEEEeqnarray}{rCl}
\label{eq: LB1, k even}
\hspace*{-0.3cm} \bigcard{ \set{P}_k } &\geq& \card{\E{G}}^k \,
\exp \bigl( -(\tfrac12 k-1) [ \Ent{P} + \Ent{Q} ] \bigr) \nonumber \\[0.1cm]
&& \cdot \exp(-\min\{ \Ent{P}, \Ent{Q}\} \bigr) \\[0.2cm]
\label{eq: LB2, k even}
&\geq& \frac{\card{\E{G}}^k}{(mn)^{\frac{k}{2}-1} \, \min\{m,n\}},
\end{IEEEeqnarray}
\end{enumerate}
with equalities in \eqref{eq: LB1, k odd}--\eqref{eq: LB2, k even} if the
bipartite graph $G$ is regular.
\end{proposition}

\begin{IEEEproof}
We prove Proposition~\ref{proposition: number of walks}
when $k \in \naturals$ is odd. The proof when
$k$ is even is essentially similar.

Let $k \eqdef 2 \ell + 1$ with $\ell \in \naturals$, and select a random walk of
length $k$ in the bipartite graph $G$ by the following procedure:
\begin{enumerate}[1)]
\item The edge $E_{\ell+1} \in \E{G}$ is selected uniformly at random among all edges in $G$.
Let $E_{\ell+1} = (U_{\ell+1}, V_{\ell+1})$ with $U_{\ell+1}$ and $V_{\ell+1}$ denoting, respectively,
the left and right vertices attached to this edge;
\item Given $E_{\ell+1}$, the edge $E_{\ell} = (U_{\ell}, V_{\ell}) \in \E{G}$ is selected uniformly
at random among all edges in $G$ with an endpoint at the given vertex $U_{\ell+1}$ (so $U_{\ell+1}=V_{\ell}$);
\item Similarly, given $E_{\ell+1}$, the edge $E_{\ell+2}$ is selected uniformly at random,
independently of $E_{\ell}$, among all edges in $G$ with an endpoint at the given vertex $V_{\ell+1}$ (so, $U_{\ell+2} = V_{\ell+1}$);
\item Given $E_{\ell}$, the edge $E_{\ell-1}$ is selected uniformly at random among all edges in $G$ with
an endpoint that coincides with the already selected endpoint $U_{\ell}$ of $E_{\ell}$ (see Step~2), and it is conditionally
independent of the edges $E_{\ell+2}$ and $E_{\ell+1}$;
\item Given $E_{\ell+2}$, the edge $E_{\ell+3}$ is selected uniformly at random among all edges in $G$ with
an endpoint that coincides with the already selected endpoint $V_{\ell+2}$ of $E_{\ell+2}$ (see Step~3), and it is conditionally
independent of $E_{\ell+1}, E_{\ell}$ and $E_{\ell-1}$;
\item The process of selecting the edges of the walk is continued this way, and the following Markov chain is constructed:
\begin{IEEEeqnarray}{rCl}
\label{eq: Markov chain}
\hspace*{-1cm} E_1 \, \markov \ldots \, \markov \, E_{\ell} \, \markov \, E_{\ell+1}
\, \markov \, E_{\ell+2} \, \ldots \, \markov \, E_{2\ell+1}.
\end{IEEEeqnarray}
\end{enumerate}
By the construction of a $k$-length random walk in $G$ as above,
\begin{IEEEeqnarray}{rCl}
\hspace*{-0.5cm} \log \card{\set{P}_k} & \geq & \Ent{E_1, \ldots, E_{2\ell+1}} \label{eq: 21012021a1} \\
\hspace*{-0.5cm} &=& \Ent{E_{\ell+1}} + \bigl[ \EntCond{E_{\ell+2}}{E_{\ell+1}} + \EntCond{E_{\ell}}{E_{\ell+1}} \bigr] \nonumber \\[0.1cm]
&& + \bigl[ \EntCond{E_{\ell+3}}{E_{\ell+2}} + \EntCond{E_{\ell-1}}{E_{\ell}} \bigr] + \ldots  \nonumber \\[0.1cm]
&& + \bigl[ \EntCond{E_{2\ell+1}}{E_{2\ell}} + \EntCond{E_1}{E_2} \bigr]  \label{eq: 21012021a2}
\end{IEEEeqnarray}
where \eqref{eq: 21012021a1} holds since $(E_1, \ldots, E_{2\ell+1})$ is a random walk in $\set{P}_k$;
\eqref{eq: 21012021a2} holds by the chain rule of the Shannon entropy
and the Markovity property in \eqref{eq: Markov chain}. By Step~1 of the construction,
\begin{IEEEeqnarray}{rCl}
\Ent{E_{\ell+1}} = \log \bigcard{\E{G}}, \label{eq: 21012021a3}
\end{IEEEeqnarray}
and, by Steps~2 and~3 of the construction,
\begin{IEEEeqnarray}{rCl}
&& \EntCond{E_{\ell+2}}{E_{\ell+1}} = \sum_{v \in V} \biggl\{ \frac{\dgr{v}}{\bigcard{\E{G}}} \cdot \log \dgr{v} \biggr\}, \label{eq: 21012021a4} \\[0.1cm]
&& \EntCond{E_{\ell}}{E_{\ell+1}} = \sum_{u \in U} \biggl\{ \frac{\dgr{u}}{\bigcard{\E{G}}} \cdot \log \dgr{u} \biggr\}. \label{eq: 21012021a5}
\end{IEEEeqnarray}
Moreover, by Steps~4--6, for all $j \in \{1, \ldots, \ell\}$,
\begin{IEEEeqnarray}{rCl}
&& \EntCond{E_{\ell+j+1}}{E_{\ell+j}} = \EntCond{E_{\ell+2}}{E_{\ell+1}}, \label{eq: 21012021a6} \\[0.1cm]
&& \EntCond{E_{\ell+1-j}}{E_{\ell+2-j}} = \EntCond{E_{\ell}}{E_{\ell+1}}.  \label{eq: 21012021a7}
\end{IEEEeqnarray}
For $k = 2 \ell + 1$ with $\ell \in \naturals$, it follows that
\begin{IEEEeqnarray}{rCl}
\hspace*{-1cm} \log \card{\set{P}_k} & \geq & \log \bigcard{\E{G}} + \ell \, \Biggl[ \,
\sum_{u \in U} \biggl\{ \frac{\dgr{u}}{\bigcard{\E{G}}} \cdot \log \dgr{u} \biggr\} \nonumber \\[0.1cm]
&& \hspace*{2cm} + \sum_{v \in V} \biggl\{ \frac{\dgr{v}}{\bigcard{\E{G}}} \cdot \log \dgr{v} \biggr\} \Biggr] \label{eq: 21012021a8} \\[0.1cm]
&=& \log \bigcard{\E{G}} + \ell \, \Biggl[ \,
\sum_{u \in U} \biggl\{ \frac{\dgr{u}}{\bigcard{\E{G}}} \cdot \log \frac{\dgr{u}}{\bigcard{\E{G}}} \biggr\} \nonumber \\[0.1cm]
&& \hspace*{2cm} + \sum_{v \in V} \biggl\{ \frac{\dgr{v}}{\bigcard{\E{G}}} \cdot \log \frac{\dgr{v}}{\bigcard{\E{G}}} \biggr\} \nonumber \\[0.1cm]
&& \hspace*{2cm} + 2 \log \bigcard{\E{G}} \Biggr]  \label{eq: 21012021a9} \\[0.1cm]
&=& (2 \ell + 1) \, \log \bigcard{\E{G}} - \ell \bigl[ \Ent{P} + \Ent{Q} \bigr] \label{eq: 21012021a10} \\[0.1cm]
&=& k \log \bigcard{\E{G}} - \tfrac12 (k-1) \bigl[ \Ent{P} + \Ent{Q} \bigr], \label{eq: 21012021a11}
\end{IEEEeqnarray}
where \eqref{eq: 21012021a8} holds by \eqref{eq: 21012021a1}--\eqref{eq: 21012021a7};
\eqref{eq: 21012021a9} holds since ($G$ is bipartite)
\begin{IEEEeqnarray}{rCl}
\sum_{u \in U} \dgr{u} = \bigcard{\E{G}} = \sum_{v \in V} \dgr{v};  \label{eq: 21012021a12}
\end{IEEEeqnarray}
\eqref{eq: 21012021a10} holds by the definition of the PMFs $\pmf{P}$ and $\pmf{Q}$ in \eqref{eq: PMF P - left side}
and \eqref{eq: PMF Q - right side}, respectively. Finally, exponentiating the left side in \eqref{eq: 21012021a9}
and the right side in \eqref{eq: 21012021a11} gives the lower bound in \eqref{eq: LB1, k odd}.

The transition from the lower bound on $\bigcard{\set{P}_k}$ in the right side of \eqref{eq: LB1, k odd}
to the looser lower bound in the right side of \eqref{eq: LB2, k odd} holds since $\card{\set{U}} = m$ and
$\card{\set{V}} = n$ yields (see \eqref{eq: PMF P - left side} and \eqref{eq: PMF Q - right side})
\begin{IEEEeqnarray}{rCl}
\Ent{P} \leq \log m, \quad \Ent{Q} \leq \log n.
\end{IEEEeqnarray}  

We finally show that the lower bound in the right side of \eqref{eq: LB2, k odd} is achieved if $G$ is a regular bipartite graph.
Let $G$ be a bipartite graph with fixed degrees
$d_{\mathrm{L}}$ and $d_{\mathrm{R}}$ on its left and right sides, respectively. Then,
$m d_{\mathrm{L}} = \bigcard{\E{G}} = n d_{\mathrm{R}}$, and
\begin{IEEEeqnarray}{rCl}
\frac{\card{\E{G}}^k}{(mn)^{\frac{k-1}{2}}}
&=& \bigcard{\E{G}} \, (d_{\mathrm{L}} \, d_{\mathrm{R}})^{\frac{k-1}{2}},  \label{eq: 21012021a15}
\end{IEEEeqnarray}
which is the number of $k$-length walks in $G$ for odd $k$.
\end{IEEEproof}

A certain non-returning walk was considered in \cite{AlonHL02} for graphs of minimum 
degree at least~2. It is left for a future study to examine the suitability of the 
same idea to yield bounds similar to Proposition~\ref{proposition: number of walks} on
the number of $k$-length trails (i.e., walks with no repeated edges), and the number of
$k$-length paths (i.e., walks without repeated edges and vertices).

\vspace*{0.2cm}
\section*{Acknowledgement} Constructive comments by the three anonymous reviewers
are gratefully acknowledged.

\end{document}